\documentclass[preprint,groupedaddress,floatfix,%
nofootinbib]{revtex4}

\usepackage{euscript}
\usepackage{dcolumn}
\usepackage{graphicx}
\usepackage{epsfig}
\usepackage{hyperref}
\usepackage{graphicx}
\usepackage{dcolumn}
\usepackage{longtable}
\usepackage{times}
\usepackage{amsmath,amssymb,bm}
\usepackage[english]{babel}
\usepackage[latin1]{inputenc}
\usepackage{times}
\usepackage[T1]{fontenc}
\usepackage{color}
\usepackage{fancybox}
\usepackage{sidecap}
\usepackage{graphicx}
\usepackage{epsfig}
\usepackage{psfrag}
\usepackage{bbm}
\usepackage[english]{babel}
\usepackage[latin1]{inputenc}
\usepackage{times}
\usepackage[T1]{fontenc}
\usepackage{color}
\usepackage{fancybox}
\usepackage{sidecap}
\usepackage{psfrag}
\usepackage{bbm}
\usepackage{wasysym}
\usepackage{slashed}
\usepackage{tikz}
\usepackage{euscript}

\begin{document}

\title{Stabilizer Approximation}
\author{Xinying Li}
\author{Jianan Wang}
\author{Chuixiong Wu}
\author{Fen Zuo\footnote{Email: \textsf{zuofen@miqroera.com}}}
\affiliation{Shanghai MiQro Era Digital Technology Ltd, Shanghai, China}

\begin{abstract}
We propose a heuristic method to obtain the approximate groundstate for a Hamiltonian in the qubit form, based on the stabilizer formalism. These states may serve as proper initial states for further refined computation. It would be interesting to assess the efficiency and scalability of the method.

\end{abstract}
 \maketitle

\tableofcontents

\section{INTRODUCTION}
Recently the stabilizer formalism~\cite{Gottesman96,Gottesman97,Gottesman98} is utilized to obtain approximate groundstates of chemical and manybody physical systems~\cite{Shang2021, CAFQA, CAFQA2}. However, these studies focus on the normalizers, namely the Clifford circuits, rather than the stabilizers themselves, and thus are not as efficient as expected~\cite{Gottesman98}. To fully capture the essence of the stabilizer formalism, it would be better to work with the stabilizers directly. For the chemical and manybody problems, they could be engineered naturally from the qubit form of the Hamiltonians. In the following we will illustrate such a procedure using some examples. In some sense, our method could be considered as a follow-up to the qubit-tapering algorithm~\cite{Tapering}.

\section{THE METHOD}
We will take the H$_2$, LiH and BeH$_2$ molecules as examples, and work in the STO-3G basis. The distance unit is chosen to be \AA, and energy unit hartree. The chemical framework PySCF~\cite{PySCF} is utilized to generate the second-quantized fermionic Hamitonians, and do the restricted/unrestricted Hartree-Fock calculations for comparison. The open source package Qiskit~\cite{Qiskit} is then employed to transform the fermonic Hamiltonians to the qubit form. We choose the parity transformation~\cite{Parity}, but the other transformations work as well. The nuclear-nuclear potentials are not included, and will be added to the final results by hand.

\subsection{H$_2$}
Using two-qubit reduction technique~\cite{Tapering}, the H$_2$ molecule can be described with 2 qubits. Let us start from the equilibrium point $d=0.74$, where we know the Hartree-Fock solution approximates the exact result well. The Hamiltonian reads
 \begin{eqnarray}
H_{\text{H}_2}&=& -1.0534210769165204 * II\nonumber\\
&&+ 0.39484436335590356 * IZ\nonumber\\
&&- 0.39484436335590367 * ZI\nonumber\\
&&+ 0.1812104620151969 * XX\nonumber\\
&&- 0.011246157150821112 * ZZ.\nonumber
\end{eqnarray}
To approach the groundstate as close as possible, the stabilizers should be chosen to make Hamiltonian as small as possible. It is straightforward to figure out the best choice is
\begin{equation}
Z_1, \quad -Z_0. \nonumber
\end{equation}
Here the qubits are labeled from right to left as $0,1,2...$. And for $M/2$ spatial orbitals, the first $M/2$ qubits correspond to the spin-up spin orbitals, and the latter $M/2$ qubits correspond to the spin-down ones. So the stabilizer state is fixed to be $|0_11_0\rangle$, which is simply the parity transformation of the corresponding Hartree-Fock state. The corresponding total energy is about $-1.12$, while the exact value is $-1.14$.

Now we proceed to study the region away from the equilibrium point. Taking $d=2.8$, the Hamiltonian reads:
\begin{eqnarray}
H'_{\text{H}_2}&=& -0.8284676561247681 * II \nonumber \\
&&+ 0.2930431286727852 * XX\nonumber \\
&&+ 0.016170000066607376 * IZ \nonumber \\
&&- 0.016170000066607328 * ZI  \nonumber \\
&&- 0.0001469354633982234 * ZZ.  \nonumber
\end{eqnarray}
It is straightforward to choose the first stabilizer as $-X_1X_0$. Then single $Z$ operators are not allowed to be stabilizers, and we have to choose $Z_1Z_0$. The stabilizer state is then
\begin{equation}
\frac{|0_10_0\rangle-|1_11_0\rangle}{\sqrt{2}}.\label{eq.en1}
\end{equation}
The corresponding total energy is about $-0.93$, nearly the same as the exact result. However, the coefficient of the $Z_1Z_0$ term is very close to zero, which indicates a possible two-fold degeneracy. If so, $-Z_1Z_0$ could be an alternative choice for the second stabilizer, and the corresponding state is
\begin{equation}
\frac{|0_11_0\rangle-|1_10_0\rangle}{\sqrt{2}}.\label{eq.en2}
\end{equation}
The splitting of the two states we see here could be an artificial effect of the qubit reduction procedure. Also we find that the particle number explanation is a little tricky in such a situation.

So let us go back to the original $4$-qubit form. The Hamiltonian at $d=2.8$ is given by
\begin{eqnarray}
H''_{\text{H}_2}&=&-0.7340910665455848 * IIII\nonumber\\
&&+ 0.12194654795642478 * ZZZZ\nonumber\\
&&+ 0.12044907695778825 * ZZIZ\nonumber\\
&&+ 0.12044907695778825 * IZIZ\nonumber\\
&&+ 0.11909854142254998 * IZZZ\nonumber\\
&&+ 0.07326078216819633 * ZXIX\nonumber\\
&&- 0.07326078216819633 * IXZX\nonumber\\
&&- 0.07326078216819633 * ZXZX\nonumber\\
&&+ 0.07326078216819633 * IXIX\nonumber\\
&&+ 0.04718829478959189 * IIZI\nonumber\\
&&+ 0.04718829478959189 * ZIZI\nonumber\\
&&+ 0.047527808451266154 * IIIZ\nonumber\\
&&+ 0.04752780845126615 * IZZI\nonumber\\
&&+ 0.03135780838465877 * IIZZ\nonumber\\
&&+ 0.03135780838465882 * ZZII.\nonumber
\end{eqnarray}
After some work one will find out the first three stabilizers could be chosen as
\begin{equation}
-Z_2Z_0, Z_1, Z_3.\nonumber
\end{equation}
 But then the Hamiltonian gives no more information for us to fix the last stabilizer. So we conclude that the stabilizer groundstate is indeed 2-fold degenerate, and could be taken to be
\begin{equation}
\frac{|0_30_20_11_0\rangle+|0_31_20_10_0\rangle}{\sqrt{2}}, \quad \frac{|0_30_20_11_0\rangle-|0_31_20_10_0\rangle}{\sqrt{2}}.\label{eq.en3}
\end{equation}
This means the last stabilizer has been chosen as $X_2X_0$ and $-X_2X_0$, respectively. However, the degeneracy actually destroys the entanglement structure in the above states, making them equivalent to product states, or Hartree-Fock states. A unrestricted Hartree-Fock calculation confirms such a conclusion.

One could sample more points on the potential energy curve. It is not difficult to show that the potential curve at small distance could be well approximated by the Hartree-Fock/product states, while at large distance is well approximated by the entangling states (\ref{eq.en1},\ref{eq.en2},\ref{eq.en3}). This is in accordance with the numerical results in~\cite{CAFQA}.

\subsection{LiH}
To avoid the degeneracy of H$_2$, we go ahead to study the asymmetric LiH molecule. One can estimate the required qubits by counting the atomic orbitals. We take the $1s, 2s, 2p$ orbitals of Li and the $1s$ orbital of H into account, so we have in total $6$ spatial orbitals, or $12$ spin orbitals. After the two-qubit reduction, we would get a $10$-qubit Hamiltonian, which contains as many as $631$ Pauli terms.

We start from the point $d=1.5$ near the equilibrium. The Hamiltonian reads:
\begin{eqnarray}
H_{\text{LiH}}&=&-5.161946304557396 * IIIIIIIIII \nonumber\\
&&+ 1.010986985827016 * IIIIIIIIIZ  \nonumber\\
&&+ 1.0109869858270164 * IIIIZIIIII \nonumber\\
&&+ 0.4145416937880457 * IIIIZIIIIZ\nonumber\\
&&- 0.3982629120981239 * IIIIIZIIII\nonumber\\
&&- 0.39826291209812403 * ZIIIIIIIII\nonumber\\
&&- 0.22878247783636188 * ZZIIIIIIII\nonumber\\
&&- 0.2287824778363618 * IZZIIIIIII\nonumber\\
&&- 0.22878247783636196 * IIIIIZZIII\nonumber\\
&&- 0.2287824778363618 * IIIIIIZZII\nonumber\\
&&- 0.19731447155414064 * IIIIIIIZZI\nonumber\\
&&- 0.19731447155414092 * IIZZIIIIII\nonumber\\
&&- 0.11495682604662383 * IIIIIIIIZZ\nonumber\\
&&- 0.11495682604662373 * IIIZZIIIII\nonumber\\
&&+ ...\nonumber
\end{eqnarray}
From these dominant terms one can easily fix the stabilizers to be
\begin{equation}
Z_9, Z_8, Z_7, Z_6, -Z_5, Z_4, Z_3, Z_2, Z_1, -Z_0. \nonumber
 \end{equation}
So the stabilizer state is the product state
\begin{equation}
|0_90_80_70_61_50_40_30_20_11_0\rangle,\label{eq.HF}
\end{equation}
or Hartree-Fock state, as expected. The total energy can be estimated to be $-7.86$, close to the exact value $-7.88$.

To see the entangling structure in the faraway region, we generate the Hamiltonian at $d=5.0$, which is given by :
\begin{eqnarray}
H'_{\text{LiH}}&=&-4.712365947275493 * IIIIIIIIII \nonumber\\
&&+ 1.0043414021528052 * IIIIIIIIIZ \nonumber\\
&&+ 1.0043414021528052 * IIIIZIIIII\nonumber\\
&&+ 0.415119697523212 * IIIIZIIIIZ\nonumber\\
&&- 0.23366481986596774 * IIIIIZIIII\nonumber\\
&&- 0.2336648198659678 * ZIIIIIIIII\nonumber\\
&&- 0.23360151362602502 * IIIIIIZZII\nonumber\\
&&- 0.23360151362602524 * IIIIIZZIII\nonumber\\
&&- 0.23360151362602505 * IZZIIIIIII\nonumber\\
&&- 0.23360151362602527 * ZZIIIIIIII\nonumber\\
&&- 0.12132590572285508 * IIZZIIIIII\nonumber\\
&&- 0.12132590572285518 * IIIIIIIZZI\nonumber\\
&&- 0.11613981631234098 * IIIIIIIIZZ\nonumber\\
&&- 0.11613981631234092 * IIIZZIIIII\nonumber\\
&& + ... \nonumber
\end{eqnarray}
From these dominant terms one can already get $8$ stabilizers:
\begin{equation}
Z_9, Z_8, Z_7, -Z_5, Z_4, Z_3, Z_2, -Z_0. \nonumber
\end{equation}
 A detailed analysis of some subdominant terms suggests the remaining two stabilizers be chosen as $-Z_6Z_1$ and $-X_6X_1$. So the resulting state is the entangling state:
\begin{equation}
\frac{|0_90_80_71_61_50_40_30_20_11_0\rangle-|0_90_80_70_61_50_40_30_21_11_0\rangle}{\sqrt{2}}. \label{eq.en4}
\end{equation}
The corresponding total energy is roughly $-7.76$, again close to the exact value $-7.78$. Now we do not have any degeneracy any more, and the entangling structure do not collapse. So in principle, the above entangling state can not be directly obtained through a Hartree-Fock procedure, which only samples the product states. However, by treating the spin-up and spin-down sectors separately as in the unrestricted Hartree-Fock approach, one could somehow imitate the entangling effect. Numerical calculation shows that unrestricted Hartree-Fock approach gives similar results for the total energy in this region.

Same as the H$_2$ molecule, one can check that the product state (\ref{eq.HF}) and the entangling state (\ref{eq.en4}) together give the best stabilizer approximation for the potential energy curve in the whole region. Again this is in accordance with the numerical results in~\cite{CAFQA}.

\subsection{BeH$_2$}

For molecules with more than two atoms, one would expect more dominant stabilizer configurations, and configurations with complicated entangling structure. So we apply the method to the BeH$_2$ molecule. Near the equilibrium point the atoms are on a line, with Be at the middle. We take the Be-H distance $d$ as the varying parameter, consider the equilibrium point $d=1.32$, and the faraway point $d=5.0$. The procedure is nearly the same as LiH, so we simply list the results.

At $d=1.32$, the stabilizers are:
\begin{equation}
Z_{11},Z_{10},Z_9, Z_8, -Z_7, Z_6, -Z_5, -Z_4, -Z_3, -Z_2, Z_1, -Z_0. \nonumber
 \end{equation}
The corresponding state is
\begin{equation}
|0_{11}0_{10}0_90_81_70_61_51_41_31_20_11_0\rangle.\label{eq.HF2}
\end{equation}
The total energy is about $-15.56$, while the exact energy is $-15.60$.

At $d=5.0$, the first $11$ stabilizers can be easily chosen as:
\begin{equation}
Z_{11}, Z_{10},Z_9, -Z_7, Z_6, -Z_5, -Z_4, -Z_3, Z_1, -Z_0, -X_8X_2. \nonumber
\end{equation}
Again here the Hamiltonian gives no more information for the last stabilizer. This is because, just as H$_2$, BeH$_2$ is also symmetric, and the groundstate is doubly degenerate. For simplicity, let us choose the last stabilizer to be $Z_8Z_2$, or $-Z_8Z_2$. The corresponding stabilizer states are
\begin{eqnarray}
&&\frac{|0_{11}0_{10}0_90_81_70_61_51_41_30_20_11_0\rangle-|0_{11}0_{10}0_91_81_70_61_51_41_31_20_11_0\rangle}{\sqrt{2}},\label{eq.en5}
\\
&&\frac{|0_{11}0_{10}0_90_81_70_61_51_41_31_20_11_0\rangle-|0_{11}0_{10}0_91_81_70_61_51_41_30_20_11_0\rangle}{\sqrt{2}}.\label{eq.en6}
\end{eqnarray}
Notice the similarity with (\ref{eq.en1}) and (\ref{eq.en2}). Again the particle number explanation is ambiguous. The total energy is about $-15.28$, to be compared with the exact value $-15.34$. The unrestricted Hartree-Fock procedure gives an energy value -$15.22$, which is close to but not exactly the same as ours.

Unfortunately, BeH$_2$ does not give rise to any new structure, probably due to the symmetry of the system.

\section{DISCUSSION}

As a generation of the traditional Hartree-Fock states, stabilizer states could approximate the true groundstates reasonably well. We find that the corresponding stabilizers could be naturally engineered from the qubit-form Hamiltonians. Importantly, a specific stabilizer configuration seems to dominate over a large region, making the calculation of the potential energy surface much easier. For the simple molecules analyzed here, the stabilizer configuration exhibits an interesting pattern, a transition from the product state/Hartree-Fock state near the equilibrium to the entangling state in the faraway region.

However, since the method is heuristic, it is not clear whether the method will always be efficient. Also it is not verified if it performs well in much larger systems.

We will continue to explore the properties of the method to see if it qualifies as a satisfactory initial-state preparation procedure, which would be important for further refined computation~\cite{GKC2208}.

\section*{CODE AVAILABILITY}
A preliminary version of the algorithm is developed at https://github.com/MiqroEra/Stabilizer.


\end{document}